# Magnetically Compensated Nanometer-thin Ga-Substituted Yttrium Iron Garnet (Ga:YIG) Films with Robust Perpendicular Magnetic Anisotropy

*Carsten Dubs,\* Oleksii Surzhenko*

Dedicated to Prof. Dr. Peter Görnert, 1943 - 2023

INNOVENT e.V. Technologieentwicklung, 07749 Jena, Germany

c.dubs@innovent-jena.de

## Abstract

Magnetically full or partially compensated insulating ferrimagnets with perpendicular magnetic anisotropy (PMA) offer valuable insights into fundamental spin-wave physics and high-speed magnonic applications. This study reports on key magnetic parameters of nanometer-thin Ga substituted yttrium iron garnet (Ga:YIG) films with saturation magnetization $4\pi M_s$ below 200 G. Vibrating sample magnetometry (VSM) is used to determine the remanent magnetization $4\pi M_r$ and the polar orientation of the magnetic easy axis in samples with very low net magnetic moments. Additionally, the temperature dependence of the net magnetization of magnetically compensated Ga:YIG films, with compensation points $T_{comp}$ near room temperature is investigated. For films with remanent magnetization values below 60 G at room temperature, the compensation points $T_{comp}$ are determined and correlated with their Curie temperatures $T_C$. Ferromagnetic resonance (FMR) measurements at 6.5 GHz show that the FMR linewidths $\Delta H_{FWHM}$ correlate inversely proportional with the remanent magnetization. The reduced saturation magnetization in the Ga:YIG films leads to a significant increase in the effective magnetization $4\pi M_{eff}$ and thus enables films with robust PMA. This opens up a new parameter space for the fine-tuning of potential magnonic spin-wave devices on commonly used GGG substrates.

# 1. Introduction

For spin-wave devices, ferro- or ferrimagnetic insulator films with perpendicular magnetic anisotropy (PMA) and low magnetic damping losses are highly desirable to achieve isotropic spin-wave transport over large distances. Investigating material systems with uniaxial anisotropy, characterized by an out-of-plane magnetic easy axis, and using fast exchange-dominated spin waves of large wavelengths that possess a strong isotropic dispersion, is promising in this context. This approach could enable the operation of magnonic waveguide microstructures where important parameters such as the wave velocity and phase accumulation are largely independent of the structure size and magnetization orientation. Nanometer-thin rare-earth iron garnet (REIG) films with rare-earth elements (RE) such as yttrium,[1-3] terbium[4], thulium,[5-7] and gadolinium[8] represent a suitable class of materials for this purpose. The films deposited on common or substituted (111) gadolinium gallium garnet (GGG) substrates exhibit large stress-induced anisotropy contributions due to lattice mismatched film/substrate combinations (see e.g. Ref.[9-10]). These lead to the desired PMA behavior. However, in the case of bismuth-substituted iron garnet films deposited on substituted GGG substrates,[11-14] growth-induced magnetic anisotropy significantly contributes to the pronounced PMA. Despite the resulting high stresses in these films, the formation of the garnet phase is feasible under non-equilibrium process conditions, because the deposited material cannot leave the substrate surface after deposition and thus heavily strained film lattices can form. However, this is unachievable in equilibrium deposition epitaxy processes such as liquid phase epitaxy (LPE), where large lattice strains require large degrees of supersaturation that cannot be achieved by supercooling, so that nanometer-thin films cannot be grown under these conditions. To achieve PMA in yttrium iron garnet (YIG) LPE films, we pursued a different approach by epitaxially growing partially compensated Ga-substituted YIG (Ga:YIG) films on common GGG substrates. For Ga:YIG, the incorporation of paramagnetic rare earths, which typically broaden the FMR linewidth (see e.g. **Table S2** in the Supporting Information), or the application of excessive lattice strain via non-lattice matched substrates is not required to obtain films with robust PMA. Here, PMA is achieved by the so called "nucleation-induced coercivity" approach[15,16] by reducing the saturation magnetization $4\pi M_s$ (and thus the demagnetizing energy $2\pi M_s^2$) to below 200-300 G. This reduction prevents the immediate nucleation of demagnetizing domains at fields below saturation. In addition, this process is supported by a moderate contribution from the stress-induced anisotropy caused by the magnetostrictive component due to the lattice misfit.[17] However, an excessive Ga content induces significant tensile stress resulting in strong mechanical tensions that causes cracking in micrometer thick LPE films on GGG substrates.[18] By maintaining the film thickness below a critical thickness, approximately 400-500 nm in our case, crack-free Ga:YIG[19-21] films can be successfully grown on GGG substrates. The potential of such substituted films for real-world magnonic devices has already been discussed by various authors. For example, Golebiewski et al.[22] reported on an analysis of field-programmable gate array (FPGA) circuits, suggesting that three times higher clock rates could be achieved with Ga:YIG films than with YIG films of the same wavenumber. Wang et al.[23] discussed the energy efficiency of Ga:YIG, and noted that it could enable energy consumption as low as 0.5 atto-joules per bit, that would be the lowest value reported for magnonic logic devices in bias-free applications. Therefore, investigating nanometer-thin films of Ga:YIG on (111) GGG substrates with reduced magnetization and bias-free perpendicular magnetization states promises new insights into fundamental spin wave physics and high-speed magnonic applications.

## 2. Phenomenological description of nonmagnetically substituted YIG

Natural and synthetic garnets are characterized by the nominal formula $\{C_3\}[A_2](D_3)O_{12}$. The complex garnet unit cell is built up of dodecahedral ions C like $Y^{3+}$, $RE^{3+}$, $Bi^{3+}$, $Ca^{2+}$, etc.; octahedral ions A like $Fe^{3+}$, $Ga^{3+}$, $Al^{3+}$, $Sc^{3+}$, $In^{3+}$, etc.; and the tetrahedral ions D like $Fe^{3+}$, $Ga^{3+}$, $Al^{3+}$, $Si^{4+}$, $Ge^{4+}$, etc., all coordinated by oxygen ions. In the case of yttrium iron garnet $\{Y_3\}[Fe_2](Fe_3)O_{12}$ nonmagnetic $Y^{3+}$ ions are on the dodecahedral lattice site; only the octahedral and tetrahedral sublattices are occupied by magnetic $Fe^{3+}$ ions, which are strongly coupled to each other via antiferromagnetic superexchange.[24,25] The substitution of nonmagnetic ions into these iron sublattices is one of the key parameters for tuning the magnetic behavior of this outstanding ferrimagnetic material. The magnetic properties of such substituted iron garnets $\{Y_3\}[Fe_{2-x}A_x](Fe_{3-y}D_y)O_{12}$ (A or D denotes nonmagnetic ions; in our case $Ga^{3+}$) can be described by a two-sublattice Néel model of ferrimagnetics,[26] where the net magnetic moments of the iron garnet can be expressed by the following relation:

$$m_s(T) = m_d(T) - m_a(T) = M_s(T) \times V_{YIG} \quad (1),$$

where $m_s$ is the net magnetic moment, $m_d$ and $m_a$ are the magnetic moments of tetrahedral d-sites ($Fe_{3-y}D_y$) and octahedral a-sites $[Fe_{2-x}A_x]$ of the iron sublattices, respectively. The correlation of the magnetic moment $m_s$ with the magnetization $M_s$ can be achieved if the sample volume $V_{YIG}$ is known. For the degrees of Ga substitution with $z = x + y < 1.3$ formula units (f.u.) used in this study, the sublattice magnetization canting can be neglected[27,28] as only minor deviations from the simple Néel model are expected. Dopants such as $Sc^{3+}$ and $In^{3+}$ preferentially substitute the $Fe^{3+}$ ions on the octahedral a-sites, whereas the $Ga^{3+}$ and $Al^{3+}$ preferentially substitute the tetrahedral d-sites.[29,30] Both types of substituents lower the Curie temperature ($T_C$), above which the material becomes paramagnetic. For Ga substitution $z \sim 1.0$ f.u. Geller et al.[28] and Czerlinsky[31] found a distribution of ~0.9 f.u. on tetrahedral d-sites and ~0.1 f.u. on octahedral a-sites. Although the formal Ga content of the samples in this study varies up to $z \sim 1.3$ f.u. (see **Figure S1** in Supporting Information) we assume a comparable distribution of Ga over the magnetic sublattices.[32] To understand how the substitution of magnetic sublattices with nonmagnetic ions changes the magnetic characteristics of Ga:YIG samples, we present the temperature dependence of the magnetization for various cases.

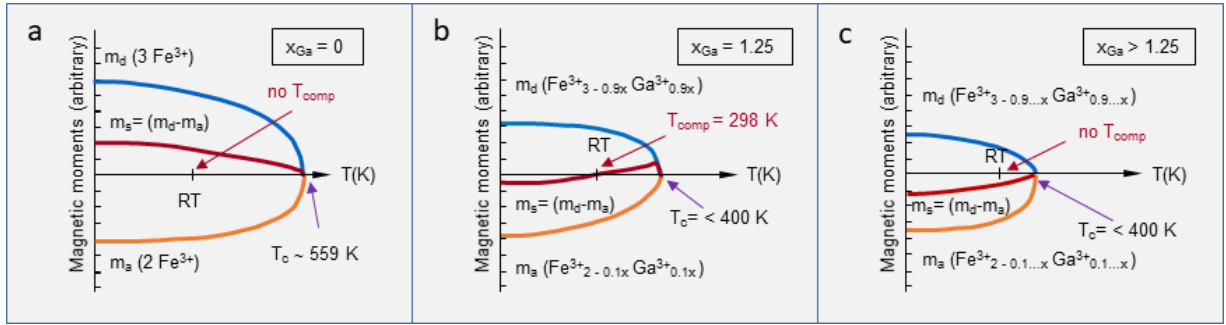

**Figure 1**

Schematic representation of the sum of the net magnetic moments $m_s$ (red curve) due to the site occupation by nonmagnetic $Ga^{3+}$ ions on the tetrahedral iron sublattice $m_d$ (blue curve) and on the octahedral iron sublattice $m_a$ (orange curve) for substituted yttrium iron garnets $Y_3Fe_{5-x}Ga_xO_{12}$ with $0 \leq x_{Ga} \leq 1.3$, as a function of temperature up to the Curie temperature $T_C$. The net magnetization, $m_s$, is dominated over the entire temperature range by the excess magnetic moments of a) the tetrahedral iron sublattice for $x_{Ga} = 0$ (no compensation points $T_{comp}$), or b) the octahedral iron sublattice below RT or the tetrahedral iron sublattice above RT for $x_{Ga} = 1.25$ ($T_{comp}$ at room temperature) or c) the octahedral iron sublattice for $x_{Ga} > 1.25$ (no compensation points $T_{comp}$). Temperature-dependent compensation points $T_{comp}$ occur when the net magnetization $m_s$ of the two antiparallel magnetic sublattices is zero ($m_s = (m_d - m_a) = 0$), while $T_C$ decreases continuously as the degree of Ga substitution increases.

**Figure 1** schematically shows the temperature-dependent contribution of the nonmagnetic $Ga^{3+}$ ions to the different magnetic sublattices $m_d$, $m_a$ and the resulting net magnetization $m_s$. In **Figure 1a**, for unsubstituted YIG, the magnetic moments $m_d$ of the tetrahedral sublattice $(Fe\uparrow\uparrow\uparrow)_d$ and the magnetic moments $m_a$ of the octahedral sublattice $[Fe\downarrow\downarrow]_a$ lead to formal net magnetic moments $m_s$ of $(Fe\uparrow)$ at 0 K. The different temperature dependency of the sublattice magnetization leads to a reduction in the net magnetization $m_s$ with increasing temperatures until the Curie temperature ($T_C$) is reached, which according to Hansen et al.[33] is 559 K for single crystalline YIG material. For pure YIG, the resulting net magnetization $m_s$ is predominantly influenced by the d-site $Fe^{3+}$ ions on the tetrahedral sublattice, which is also true for moderately Ga-substituted YIG with $z \leq 0.9$ f.u.[33,34] Beyond this level of substitution, the net magnetization approaches zero as the magnetization of the two sublattices converges. Theoretical curves calculated by Görnert et al.[34] within the molecular field model showed a crossover point at 280 K for $z = 1.26$ f.u. where the saturation magnetization is zero, referred to as compensation temperature $T_{comp}$, concurrently Curie temperature $T_C$ shifts to temperatures below 400 K. This borderline case calculated for the occupation of the different magnetic sublattices according to the Ga distribution ratio of (0.9)/[0.1] is schematically shown in **Figure 1b**. For a total Ga content of $z = 1.25$ f.u. and an occupation of the magnetic sublattices with $(Fe_{1.875}\uparrow Ga_{1.125})_d$ on tetrahedral and $[Fe_{1.875}\downarrow Ga_{0.125}]_a$ on octahedral sites, where the number of magnetic moments of the $Fe^{3+}$ ions is equal but antiparallel for both sublattices, the remanent magnetization disappears at room temperature. For this scenario samples with $z < 1.25$ f.u. have $T_{comp}$ below room temperature or no compensation point at all and $T_C$ is above 400 K.[33,34] Note that the net saturation magnetization is conventionally considered positive, but above and below the compensation point, the sublattice magnetization can have opposite signs with respect to the net magnetization $m_s$ (cf. Figure 1b). In the borderline case in **Figure 1c**, the dominant magnetization $m_a$ of the a-site $Fe^{3+}$ ions on the octahedral sublattice leads to a fully reoriented net magnetization $m_s$ with an opposite sign at room temperature and no compensation point. Therefore, for samples with z > 1.25 f.u. $T_{comp}$ shifts further towards increasing temperatures and finally coincides with $T_C$ at a temperature below 400 K (cf. Figure 1c).

## 3. Experiment and results

To investigate the magnetic and microwave properties of nonmagnetically substituted iron garnet samples, we grew epitaxial Ga:YIG films from high-temperature solutions with varying $Ga_2O_3$ content, similar to single crystal growth experiments.[33,34] Additionally, we varied the growth temperatures and supersaturations (see *Film growth* in the Experimental section). This approach allowed us to prepare LPE films with a wide range of Ga contents, achieving partial or full magnetic compensation near room temperature. To ensure that the dependence of the magnetization on Ga content corresponds to the known behavior for bulk single-crystalline samples,[33,34] we performed quantitative chemical analysis on micrometer-thick reference films grown under the same conditions as the submicrometer-thin films (see Figure S1 and the section Wavelengths-Dispersive Electron Probe Micro Analysis (WD-EPMA) measurements in Supporting Information). Due to analysis limitations for submicrometer-thin films we used the magnetization values of our submicrometer-thin films instead of the Ga content to compare our samples and correlate them with other magnetic and FMR properties in this study (see **Table S1** in Supporting Information).

### 3. 1. Magnetic Characterization

Vibration sample magnetometry (VSM) allows us to measure the net magnetic moment $m_{VSM}(H,T) = m_{Ga:YIG}(H,T) + m_{GGG}(H,T)$ of vibrating samples. In our case, these samples consist of a ferrimagnetic Ga:YIG film and a paramagnetic GGG substrate (see *Vibrating Sample Magnetometry (VSM) measurements* in the Experimental Section). Unlike other methods, VSM does not require special sample preparation and has the advantage of precisely determining the saturation magnetization $4\pi M_s(T)$, the remanent magnetization $4\pi M_r(T)$, the coercive field $H_c(T)$ and saturation field $H_s(T)$ for samples with small magnetic moments. However, a notable disadvantage is the need to extract the smallest useful signals $m_{Ga:YIG}(H,T) \ll m_{VSM}(H,T)$ for ferrimagnetic thin-film/paramagnetic substrate systems. To minimize or, ideally, to avoid the influence of the substrate signal and to accurately determine the magnetization and characteristic temperatures of the films, isothermal and temperature-dependent measurements should be conducted at small or ideally without external magnetic field. We have therefore carried out zero-field measurements, which we will discuss below.

*3.1.1. Isothermal hysteresis loops*

We tested a similar approach as described in Ref.[35] and detailed in the Supporting Information (Magnetic characterization by Vibrating Sample Magnetometry (VSM)) for Ga:YIG thin films magnetized along its easy axis (out-of-plane in our case). This approach was validated through gradual etching of a typical 'crack-free' Ga:YIG film from its as-grown thickness of $t_{Ga:YIG}$ = 445 nm down to $t_{Ga:YIG}$ = 0 nm (see **Figure 2**). The latter corresponds to a 'pure' GGG substrate whose magnetic moment $m_{GGG}$(H,T=300K) (full symbols in **Figure 2a**) appear in excellent agreement with the GGG contributions (dashed line in Figure 2a) obtained, using the aforementioned method from the $m_{VSM}$(H,T=300K) data for Ga:YIG/GGG samples of varying thicknesses. Moreover, the extracted Ga:YIG moments $m_{GaYIG}$(H,T=300K) after their normalization to the film volume $V_{Ga:YIG}$, which corresponds to the magnetization $M_{Ga:YIG}$(H,T=300K) of the respective films, are within the experimental error independent of the film thicknesses (see **Figure 2b**). This consistency confirms both the applicability of this non-destructive method and the homogeneity of the Ga:YIG material over the entire film thickness of $t_{Ga:YIG}$ = 445 nm.

While in most as-grown samples there is an abrupt switching in magnetization direction with increasing field, the observed steps in the 230 nm Ga:YIG film in Figure 2b are likely due to relieved nucleation of demagnetization domains at surface defects caused by handling during etching and following-up measurements. The reason for these steps seems to be the local reduction of the film lattice misfit at tiny scratches, which leads to a reduction of the local contribution to the uniaxial anisotropy and thus to a shift of the saturation curve to smaller values in these areas. In contrast, the magnetization of the remaining subdomains in defect-free surface areas only switches completely to the field direction when their nucleation energy is achieved.

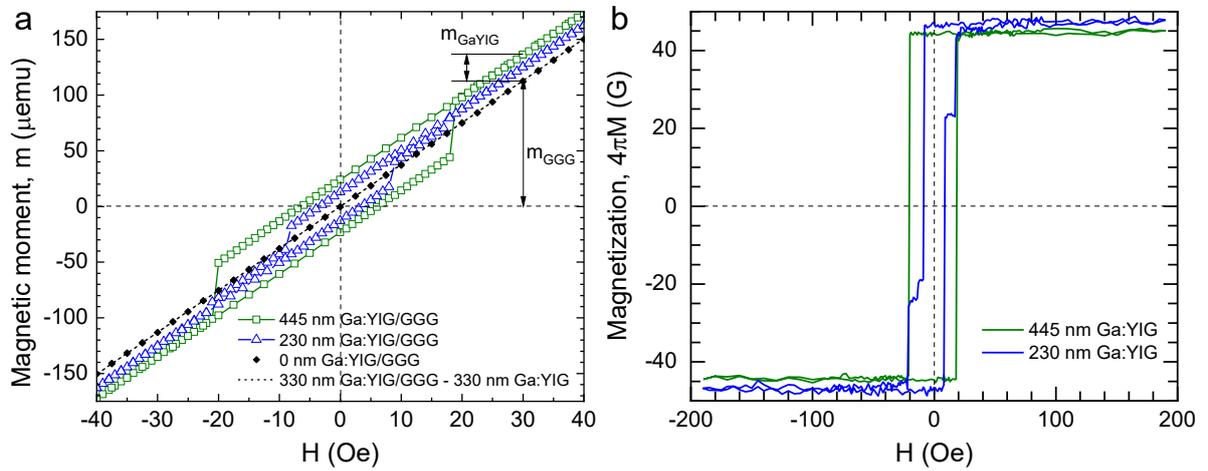

**Figure 2**

Isothermal VSM measurements of the hysteresis loops m (H,T=Const) of a typical "crack-free" Ga:YIG film after stepwise etching from an initial thickness of $t_{Ga:YIG}$ = 445 nm to $t_{Ga:YIG}$ = 0 nm. a) The curve $t_{Ga:YIG}$ = 0 nm corresponds to a completely etched Ga:YIG film ("pure" GGG substrate). The out-of-plane measured values for the magnetic moment $m_{GGG}$ (H,T=300K) (solid symbols) appear in excellent agreement with the interpolated GGG magnetic moments obtained by a mathematical extraction from the $m_{VSM}$ (H,T=300K) data e.g. of the 330 nm Ga:YIG/GGG sample (dashed line) according to the method described in Ref. [35]. b) Hysteresis loops after extracting the Ga:YIG moments and normalization to the film volume $V_{Ga:YIG}$. The magnetization values $M_{Ga:YIG}$(H,T=300K) of the as-grown and etched films from Figure 2a are the same within the experimental error.

In order to experimentally avoid the GGG substrate signal $m_{GGG}(H,T)$, and thus reduce its large influence on the Ga:YIG parameter accuracy, one should use the loop data $m_{VSM}(H,T)$ in small magnetic fields – preferably at H=0 where $m_{GGG}(H=0,T)=0$. Thus, among the key parameters derived from hysteresis loops, the remanent magnetization $M_{Ga:YIG}(H=0,T)$ is the most reliable. For this reason, we use the remanent magnetization in the following discussion to systematize the further properties of Ga:YIG films. Furthermore, owing to the square-like shape ($M_s \sim M_r$) of hysteresis loops in thin (111) Ga:YIG films with robust PMA (see Figure 2b), the remanence $M_{Ga:YIG}(H=0,T)$ should closely approximate the magnetization $M_{Ga:YIG}(H>H_s,T)$ corresponding to saturated Ga:YIG material. Therefore, zero-field VSM measurements, providing higher $m_{Ga:YIG}(H=0,T)$ accuracy, were used to define the easy axis direction in polar angle-dependent measurements, and to determine temperature dependencies $M_{GaYIG}(H=0,T)$ for thin Ga:YIG films with various magnetizations, thereby determining corresponding values for $T_C$ and $T_{comp}$.

*3.1.2. Remanence vs orientation angle measurements*

To demonstrate that the easy magnetization axis is perpendicular to the (111) Ga:YIG film surface, we measured the remanent magnetization $M_{Ga:YIG}(H=0,T=300K)$ versus the sample orientation. The results shown in **Figure 3**, confirm that the remanence in Ga:YIG films does exhibit the maxima when the magnetic field is directed perpendicular to the Ga:YIG film surface (i.e. at angles $\pm\pi n$, n=0,1,2,...).

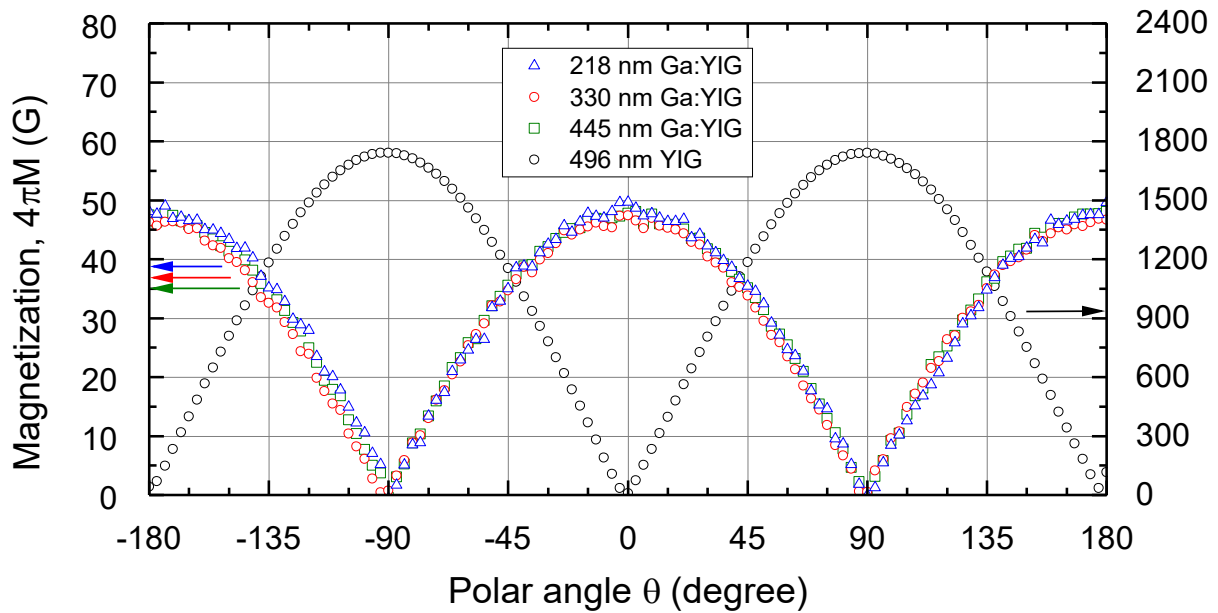

**Figure 3**

Angular dependence of the magnetization 4πM for a stepwise etched Ga:YIG film (colored symbols) and a reference YIG film (black symbols) grown on (111) GGG substrates. The 0° angle corresponds to an external magnetic field perpendicular to the film plane, while the 90° angle to a field within the plane. Note that the YIG film has an approximately 35-fold larger magnetization (right y-axis) compared to the Ga:YIG film samples (left y-axis).

In contrast, conventional YIG films are known to have their easy magnetization axis at π/2±πn (where n=0,1,2,3). Furthermore, Figure 3 illustrates that the saturation magnetization values for this Ga:YIG film ($4\pi M_r$ ~47 G) differ from these in conventional YIG material ($4\pi M_s$ ~1770 G) at room temperature by a factor of ~35 which explains the enlarged data scattering for Ga:YIG samples.

*3.1.3. Temperature-dependent remanence measurements*

As phenomenologically described above, the temperature dependence of net magnetization in nonmagnetically substituted YIG films (see Figure 1) can be used to study the sample behavior near room temperature and to determine their compensation and Curie temperatures. These measurements are carried out for different series of Ga:YIG films. **Figure 4** presents the measured temperature dependence of the remanent magnetization $4\pi M_r(H=0,T)$ for selected Ga:YIG films, with thicknesses between 100 nm and 230 nm. The corresponding quantitative values such as the film thickness t, the remanent magnetization $4\pi M_r$, the coercivity $H_c$, the compensation temperature $T_{comp}$, the Curie temperature $T_C$, the temperature difference ($T_C$-$T_{comp}$), and the FMR linewidth at 6.5 GHz are given in Table S1 in the Supporting Information.

The curves in Figure 4 show that samples #V through #Q (starting on the right side of the diagram) do not exhibit a compensation point, while samples #N through #E exhibit a compensation temperature $T_{comp}$ (cf. Figure 1b) within the studied temperature range. Sample #A seems to have only a Curie temperature (cf. Figure 1c).

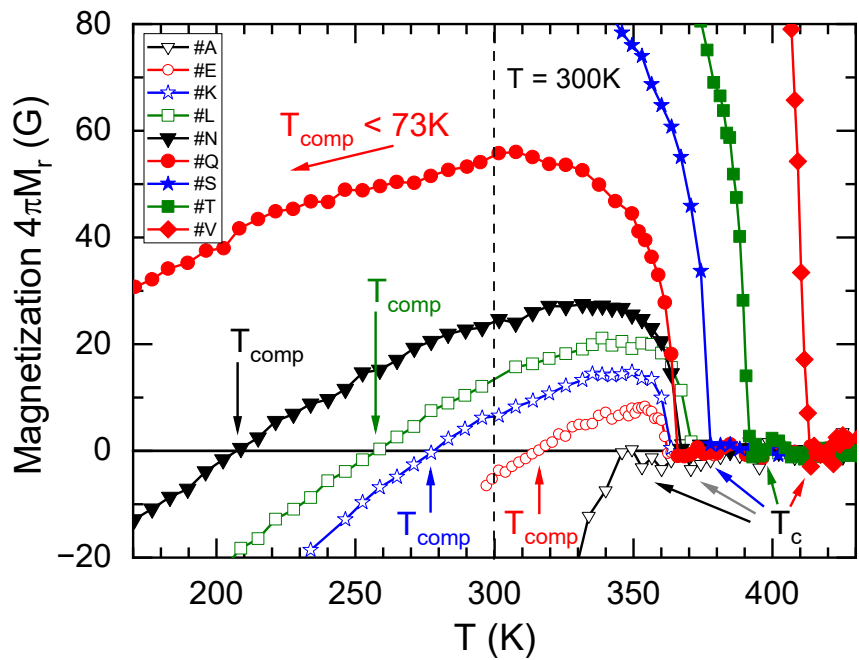

**Figure 4**

Temperature dependence of the remanent magnetization $4\pi M_r$ of Ga:YIG films with and without compensation temperatures $T_{comp}$, determined by out-of-plane VSM measurements.

With the exception of sample #Q, which has an extrapolated $T_{comp}$ value of approximately 73 K, samples #N - #E exhibit an experimentally determined $T_{comp}$ value above 200 K. This means that for nm-thin Ga:YIG films with $4\pi M_r(H=0,T=300K) \leq 60$ G magnetic compensation points exist, which can be adjusted by suitable film growth parameters to prepare Ga:YIG films with $T_{comp}$ values near room temperature T=300K. The observed temperature dependence of magnetization in Figure 4 is consistent with the experimental and theoretical predictions for single-crystalline Ga:YIG bulk material as reported by Görnert et al.[34] and Hansen et al.[33] and the systematic course of the magnetization curves confirms the predictions of the two-sublattice model of ferrimagnets for these thin-film samples. Since the incorporation of nonmagnetic ions into ferromagnetic oxides weakens the overall interaction of the exchange forces, this leads to a corresponding decrease in the Curie temperature[24]. This can be clearly seen in Figure 4 for samples #V - #Q, while for samples #N - #A with only slightly changed Ga content the determined $T_C$ values change only minimally within the error range of the temperature measurement of the VSM measurement (cf. Table S1 in the Supporting Information). However, for samples with further increased $T_{comp}$, this point can no longer be determined with certainty, as the magnetization $4\pi M_r$ ($T_{comp}<T<T_C$) lies within the measurement error. This is the case for sample #A, where no compensation point is observed, suggesting that the octahedral magnetic sublattice sites ($m_a$) dominates the magnetization of the sample up to Curie temperature $T_C$ (cf. Figure 1c).

*3.1.4. Correlation of key magnetic parameters*

To correlate the magnetic parameters, samples from different growth runs were investigated in the range of $-24$ G $\leq 4\pi M_r(H=0,T=300K) \leq +230$ G (see Table S1 in Supporting Information). To ensure that the lattice site occupancy of the nonmagnetic $Ga^{3+}$ ions reached a defined equilibrium state across the different magnetic sublattices, as described in previous studies,[32,36] the thermal history of the last process step before the measurements was standardized (see *Thermal conditioning* in the Experimental Section). This standardization ensures consistency across the samples studied.

First, we determined the dependence of the compensation temperature $T_{comp}$ on $4\pi M_r(H=0,T=300K)$ for samples that exhibited a compensation temperature as shown in **Figure 5**. With decreasing remanent magnetization of the Ga:YIG films (due to further increasing Ga substitution), the $T_{comp}$ increases continuously from 0 K to approximately 360 K. At room temperature T = 300 K, suitably substituted films reach their full magnetic compensation ($4\pi M_r = 0$ G) and for films with a further increased degree of Ga substitution, their net magnetization increases again but with the opposite sign.

The data points were interpolated using a polynomial fit curve excluding sample #A from the fit. From this fit, an interpolated value of $T_{comp} \sim 298$ K was obtained for samples with zero magnetization, confirming the accuracy of our temperature calibration, as these measurements were performed at 300 K.

For samples with values of $4\pi M_r(H=0,T=300K) > +70$ G, no compensation point exists. It appears that for samples with $4\pi M_r(H=0,T=300K) < -24$ G (see sample #A), compensation temperature $T_{comp}$ cannot be determined with confidence. To verify this, the Curie temperature values of the samples were also plotted on the graph. The fitted curves intersect at -24 G and 359 K, where $T_{comp}$ coincides with $T_C$. Beyond this point, $T_{comp}$ does not exist. Consequently, for sample #A, which lies close to this intersection, a compensation temperature $T_{comp}$ cannot be determined with certainty, and only a Curie temperature $T_C$ of approximately 348 K can be detected.

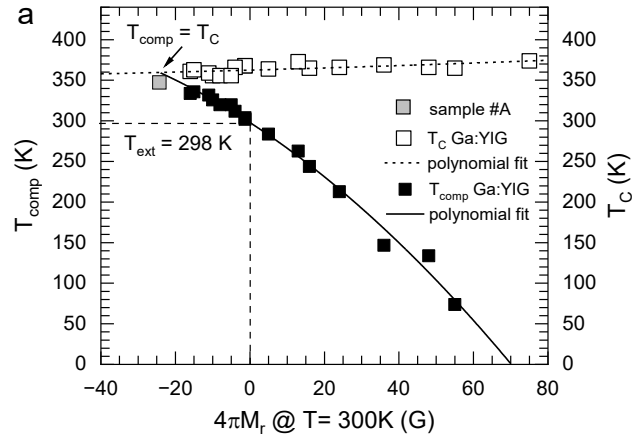

**Figure 5**

Correlation of the compensation temperature with the remanent magnetization $4\pi M_r$ (solid symbols) for Ga:YIG films thermally equilibrated at 873 K. The $T_C$ values of the samples were also plotted in the diagram (empty symbols). To guide the eyes, the data points were interpolated by polynomial fitting curves.

## 4. Microwave characterization

It is known from the literature that in Ga-substituted YIG samples, damping parameters such as the FMR linewidth $\Delta H$ and the Gilbert damping coefficient $\alpha$, which describe the FMR losses, increase with degree of substitution.[37-39] For FMR resonance measurements at 9.2 GHz and 295 K, Hansen et al.[38] reported linewidths of $\Delta H_{FWHM} = 4.5$ Oe, and $\Delta H_{FWHM} \sim 100$ Oe for single crystal spheres with saturation magnetization of $4\pi M_s = 312$ G and $4\pi M_s = 60$ G, respectively. The first value corresponds well with the values obtained for our single crystalline films with $\Delta H_{FWHM} \sim 5$ Oe for $4\pi M_r = 230$ G at 6.5 GHz and 300 K, but the second one shows a smaller $\Delta H_{FWHM}$ value of ~30 Oe at 6.5 GHz for $4\pi M_r = 55$ G (see sample #V and #Q in Table S1 in Supporting Information). The reason for this difference in the linewidths at around 60 G appears to arise from the total magnetization states of the samples. In our case, the sample's total magnetization is positive ($T_{comp} < T=300K$), whereas, in the sphere samples, the magnetization appears to be negative ($T_{comp} > T=300K$) due to a higher Ga content compared to our film sample. Unfortunately, we were unable to determine FMR linewidths for samples with $4\pi M_r < 55$ G, due to the increasing broadening of the FMR resonance peaks and the low signal intensity of the nm-thin films.

Unlike previous reports on partially magnetically compensated iron garnet single crystalline spheres,[38,39] where data sets for samples with saturation magnetization were only available until 250 G (except single values), our data set starts at this value and extends down to 0 G and beyond. In the aforementioned reports, the authors systematize their data in a normalized form using the Landau-Lifshitz damping factor[38] or the FMR linewidth[39] versus the reciprocal of the respective saturation magnetization.

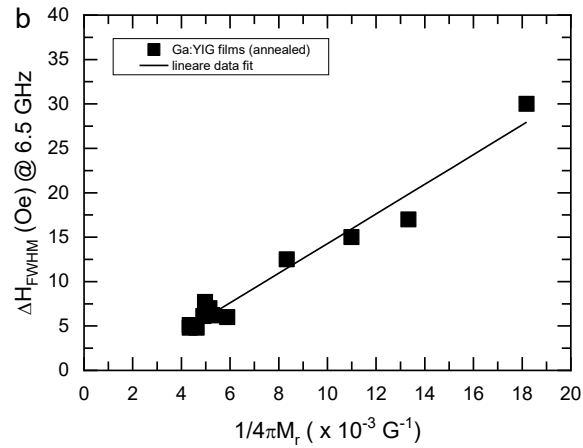

**Figure 6**

Full width of the half maximum of the FMR linewidths $\Delta H_{FWHM}$ (@ 6.5 GHz) as a function of the reciprocal remanent magnetization $1/4\pi M_r$ for Ga:YIG films with $4\pi M_r$ (T = 300K) values ≤ 250 G.

They found a linear correlation as a function of $1/4\pi M_s$ for polished single crystal spheres down to a minimum saturation magnetization of about 250 G ($1/4\pi M_s \leq 4 \times 10^{-3}$ G$^{-1}$). For the single crystalline Ga:YIG films investigated, this correlation appears to extend beyond values of $1/4\pi M_r \geq 4 \times 10^{-3}$ G$^{-1}$, as shown in **Figure 6**. However, due to the difference in the investigated saturation magnetization values and the small number of data points for our films, a direct comparison of the slopes of the linear extrapolation curves with those reported by authors[37-39] is not feasible. Nonetheless, the slope of our curves lies between these two data sets. Therefore, the influence of the GGG substrate on the film properties at room temperature does not appear significant for this correlation, as our results are comparable with measurements on spheres. This indicates that increasing Ga concentrations in the bulk material as well as in the epitaxial films leads to locally inhomogeneous, intrinsic magnetic field distributions, which cause inhomogeneous line broadening of the FMR resonance lines. Such field inhomogeneities could arise from crystalline anisotropy, inhomogeneity of the local saturation magnetization, elastic stress contributions, etc. Since the investigation of the origin of these mechanisms is beyond the scope of this study, we refer to previous work on the detailed investigation of the damping behavior of Ga:YIG samples.[38,40] The authors described the effect of inhomogeneous stresses and intrinsic relaxation contributions in relation with the thermomagnetic properties. Theoretical predictions were also presented by Widom et al.[41] who used experimental values for the magnetoelastic coefficients and the sound-wave damping crystal viscosity.

## 5. Discussion

In contrast to unsubstituted or low-substituted YIG films which typically exhibit an in-plane magnetic anisotropy, the Ga:YIG films presented in this study exhibit robust perpendicular anisotropy with intrinsic remanent magnetization. Notably, our samples exhibit total magnetization values that are at least one order of magnitude lower than previously reported for other compensated ferrimagnets, except those fully substituted with Tb or Gd.[4,8] The observed perpendicular uniaxial anisotropy contributions in the Ga:YIG films with $4\pi M_r(T=300K) \leq 200$ G arise from a combination of significantly reduced saturation magnetization (resulting in a weak demagnetization contribution) and a tensile stress-induced anisotropy due to a moderately increased lattice misfit. The formation of robust rectangular hysteresis loops is attributed to the so called "nucleation-induced" domain switching process, which is prevalent in iron garnet films with saturation magnetization below 300 G.[16] In these films, the formation of typical stripe domains after complete saturation is energetically unfavorable due to the low demagnetization energy relative to the energy required to form perpendicularly oriented domain walls. Therefore, the single-domain state persists even beyond the zero saturation field. Only when the formation of reversal domains is energetically favorable, the film magnetization reverses completely due to the stronger reversal saturation field that has already been applied. This means that such material can only be made to switch between two saturated single-domain states of opposite sign.

The modelling of magnetization reversal in (111) oriented micrometer-thick Ga:YIG films with $4\pi M_s(T=295K) = 460$ G under external magnetic fields applied perpendicular to the film direction was previously discussed by Ubizskii,[17] using experimental data from references.[33,37] In his study, the uniaxial anisotropies, resulting from the magnetostrictive components of the film due to the perpendicular lattice misfit $\Delta a^\perp$, were varied numerically. The authors calculated hysteresis curves for film parameters with non-zero lattice misfits that generate different uniaxial anisotropy $K_u$ due to defined magnetostrictive contributions. For in-plane tensile stresses caused by increasing lattice misfits with increasing Ga content, the uniaxial anisotropy contributions lead to a shift of the saturation fields to lower values. If the ratio $K_u/2\pi M_s^2$ (the uniaxial anisotropy $K_u$ ($K_{2\perp}$) relative to the demagnetization energy $2\pi M_s^2$) becomes sufficiently large, the [111] direction becomes the magnetic easy axis.[17] However, only the influence of cubic anisotropy and uniaxial anisotropy fields for a fixed saturation magnetization was considered in the modeling, without investigating the influence of decreasing film magnetization due to increasing Ga contents. Their effect can be understood by considering the different contributions to the effective perpendicular uniaxial anisotropy field $H_{2\perp}$, which is given by (cf. Supporting Information of Ref. [42])

$$H_{2\perp} = -\frac{4}{3}\frac{K_4}{M_s} + \frac{2K_\sigma}{M_s} \qquad (2).$$

For Ga:YIG films, the first term, which contains the first-order cubic anisotropy $K_4$, remains approximately constant for pure YIG and samples with varying degrees of Ga substitution. This is because $K_4$ decreases in proportion to the magnetization with increasing Ga content (see e.g. Ref. [37]), thus keeping their ratio roughly a constant. However, the second term increases because the saturation magnetization $M_s$ in the denominator decreases, while the stress-induces anisotropy $K_\sigma$ remains almost constant with increasing Ga content. The reason for the constancy of $K_\sigma$ is explained in detail in the Supporting Information (see Stress-induced anisotropy parameter $K_\sigma$).

For REIG films with substituted rare earth ions on the dodecahedral lattice sites, the same dependencies are valid. However, it is not a significantly reduced $M_s$ value at room temperature (except for Gd or Tb substitution) in the denominator of the second term of Equation 2 that increases the effective perpendicular uniaxial anisotropy field $H_{2\perp}$, but rather the value of the stress-induced anisotropy $K_\sigma$ in the numerator. Thus, the most efficient way to significantly increase the effective magnetization $4\pi M_{eff}$ (which defines the strength of the bias field required to turn the magnetization away from its magnetic easy axis direction) is to reduce $M_s$, as described by (see also Supporting Information of Ref. [42]),

$$H_{eff} = 4\pi M_{eff} = 4\pi M_s - H_{2\perp} \qquad (3).$$

In the case of Ga:YIG films, the difference between a significantly reduced $4\pi M_s$ and a large positive effective perpendicular uniaxial anisotropy field $H_{2\perp}$ leads to a strongly increased negative effective magnetization $4\pi M_{eff}$ with values exceeding -2000 G[43].

In order to evaluate the robustness and microwave applicability of nonmagnetically substituted iron garnet films such as Ga:YIG, their key parameters are listed in Table S2 in the Supporting Information in comparison to differently substituted YIG or rare earth substituted iron garnet films and discussed in detail in the Supporting Information (Benchmarking discussion). The benchmarking shows that the main advantages of the non-magnetically substituted Ga:YIG films are that they have i) a robust perpendicular magnetic anisotropy (Mr/Ms ~0.9-1.0), ii) very large effective magnetization values of $4\pi M_{eff} \geq -1500$ G, iii) the narrowest observed FMR linewidths of about 10-20 Oe @ 7 GHz, and iv) some of the lowest Gilbert damping values, regardless of their film thickness.

Thus, in addition to the usually varied parameters such as operating field $H_0$ or film thickness t, other relevant parameters such as $M_{eff}$ and $M_s$ can be used to modify the spin-wave dispersion relations of YIG. Böttcher *et al.*[20] detected for example exchange-dominated spin waves with a relatively small wave vector of k ~ 4 rad/μm for partially compensated Ga:YIG films, exhibiting a significantly more isotropic dispersion relation than dipolar waves of the same wavelength in YIG. Furthermore, it has been demonstrated that exchange-dominated spin waves, excited by either a high driving power stripline field[19] or CPW antennas,[44] can produce tunable nonlinear positive in-plane frequency shifts, with an inversion of the frequency direction compared to pure YIG. The observed nonlinear behavior also depends on the direction of the power sweep, resulting in a bistable excitation regime that generates two stable states for one given input power signal, as reported by Wang et al.[45] for YIG films.

Therefore, potential applications for these films are nonlinear spin-wave devices for bistability-based magnon computing such as logic gates, switches, amplifiers and repeaters,[46,47] as well as opportunities for neuromorphic computing devices.[48] Additionally, this PMA material allows for spin-wave propagation studies in the forward volume geometry and opens up access to fast and isotropic exchange spin wave operations.[49] This has advantages for data processing in 2D spin-wave logic devices free from non-reciprocity effects, as numerically demonstrated by Klingler et al.[50] However, the advantage of PMA films for spin-wave applications comes with challenges related to their damping behavior and the observed tendency of FMR linewidth broadening with decreasing remanent magnetization. This is likely due to additional damping process channels in Ga-substituted iron garnets, which must be taken in to account. Addressing this issue is crucial for the development of ferromagnetic resonance devices with acceptable FMR linewidths, requiring further investigation.

## 6. Conclusion

In summary, we have characterized magnetically compensated nanometer-thin Ga-substituted yttrium iron garnet films with thicknesses ranging from 90 nm to 500 nm grown by liquid phase epitaxy on GGG substrates exhibiting robust perpendicular magnetic anisotropy. Using the VSM technique, we determined the remanent film magnetization from isothermal hysteresis loop measurements, as well as the compensation temperature and Curie temperature from temperature-dependent remanence measurements. We demonstrated correlations between these characteristic temperatures and the remanent magnetization, as well as between the FMR linewidths and the reciprocal remanent magnetization. The observed PMA in our single crystalline films is primarily due to the interplay between a significantly reduced saturation magnetization and a tensile stress-induced anisotropy contribution resulting from a moderately increased lattice misfit. Therefore, the substitution of nonmagnetic $Ga^{3+}$ ions into the magnetic sublattices of iron garnets is an efficient way to significantly increase the effective magnetization of epitaxial YIG films. This opens up a new parameter space for the fine-tuning of potential magnonic spin-wave devices on commonly used GGG substrates.

## 7. Experimental Section

*Film growth:* Nanometer-thin Ga:YIG films were deposited on 1-inch (111) GGG substrates by liquid phase epitaxy from high-temperature $PbO$-$B_2O_3$-based solutions, with the addition of different amounts of $Ga_2O_3$ using the isothermal dipping method[51] . To achieve Ga:YIG films with thicknesses between 100 nm and 200 nm and a suitable Ga content, the deposition times typically ranged from one to two minutes, and deposition temperatures were between 790°C and 870°C. For the thinnest films below 100 nm, deposition times were reduced to 20 seconds, while films thicker than 200 nm required deposition times of four minutes or more.

To ensure optimal chemical homogeneity for high degrees of Ga substitution, films were deposited on horizontally rotating substrates at a speed of 200 rpm (except for a few films with remanent saturation values above 200 G, which were rotated at 100 rpm). After deposition, the samples were pulled out from the high-temperature solution, and most liquid melt residues were spun off at rotation speeds up to 1000 rpm. The samples were then immediateley pulled out of the furnace and cooled to room temperature. Subsequently, the sample holder with the sample was stored in a diluted, hot nitric-acetic-acid solution to remove any remaining solidified solution residues. Finally, the YIG film on the back of the samples was removed by mechanical polishing, and the samples were cut into chips of various sizes using a diamond wire saw.

*Thermal conditioning:* Due to the different growth temperatures, all samples were annealed in air at 873 K for 4 to 7 hours and quenched from this temperature to 673 K at a rate of about 1 K/s, followed by further cooling to room temperature according to the heat capacity of the furnace. This allows quenching of the Ga ion distribution on the magnetic sublattices and avoids a temperature-dependent and inhomogeneous Ga distribution over the sample volume. We found that an annealing time of 2 to 4 hours (depending on the film thickness) was sufficient to achieve a magnetic equlibrium state in YIG films with thicknesses between 100 – 200 nm, ensuring no significant changes in the shapes of the hysteresis loops or remanence values were observed thereafter. This ensures that thin film samples with different Ga contents may be compared accurately, despite their varying growth histories.

*Wavelengths-Dispersive Electron Probe Micro Analysis (WD-EPMA):* EPMA measurements were conducted fully quantitatively using a microprobe JXA8800L (JEOL Ltd., Tokyo, JAPAN), equipped with 4 simultaneously operating wavelengths dispersive Johansson-type crystal spectrometers. All samples were coated with approximately 5 nm carbon to avoid charging effect and were fixed on a sample holder with conductive silver paste. Due to the use of GGG substrates and the presence of Ga in both the substrate and the Ga:YIG film, it was crucial to avoid substrate excitation. To achieve this, Monte Carlo simulations were performed, and the electron energy, $E_0$, was limited to 10 keV. The Gd-$M_5N_7$ ($\alpha_1$, 1185 eV) line as one of the possible substrate signals was used as control marker to ensure that no substrate influence distorted the quantitative Ga measurement. Additionally, unsubstituted YIG films without Ga were investigated, confirming that no substrate related Ga contributions were detected at 10 keV in films with thicknesses of 2 μm. Most measurements used a probe current of 200 nA with a beam diameter defocused to 50 μm (scan off). For micrometer –thick samples exhibiting a visible network of cracks, a diameter of 20 μm was used to avoid incorporating the cracks.

*Vibrating Sample Magnetometer (VSM) measurements:* The magnetic properties of the nm-thick Ga:YIG films, grown on paramagnetic GGG substrates approximately 0.5 mm thick, were analyzed using a vibrating sample magnetometer (VSM, MicroSense LLC EZ9 equipped with the EZ1-LNA gas flow temperature control unit). Measurements were mainly conducted between 150 K and 430 K, exceeding the Curie temperature $T_C$ of the studied Ga:YIG films. The Ga:YIG/GGG samples were placed in the center of pick-up coils (with a noise ≤0.5 μemu) and aligned so that magnetic field direction was perpendicular to the film surface (out-of-plane). The net registered magnetic moment

$m(H,T) = m_{Ga:YIG}(H,T) + m_{backgr}(H,T) = M_{Ga:YIG}(H,T)V_{Ga:YIG} + [M_{GGG}(H,T)HV_{GGG} + \ldots]$ consists of the useful signal $m_{Ga:YIG}$ from the Ga:YIG sample and a background signal (due to the paramagnetic GGG substrate, diamagnetic sample holder, etc). The $m_{backgr}(H,T)$ was found to increase linearly with the external field H. This allowed us to calculate $m_{backgr\ GGG}(H,T)$ at magnetic fields high enough to saturate the Ga:YIG film and then mathematically substract background contribution over the entire H range (see e.g. Ref. [43]).

For temperature-dependent measurements of m(H=0,T), the samples were cooled, magnetized briefly in large magnetic field (~5KOe) to ensure saturation, then the field was turned off and the m(H=0,T) dependency was recorded while heating the sample until T≫$T_C$. To confirm the Curie temperatures of our Ga:YIG films, we compared VSM measurements with dynamic scanning calorimetry (DSC) measurements for a commercially available MnZn ferrite bead sample (11-685-L-1, Ferroics Inc. Bethlehem, USA), calibrated with an indium standard. The $T_C$ determination (peak-onset) of the reference sample showed good agreement between the $T_C$ ~ 368 K for the DSC analysis and the $T_C$ ~ 369 – 370 K for the VSM measurement.

*Ferromagnetic Resonance (FMR) measurements:* The FMR absorption vs frequency spectra $S_{21}(f)$ for the studied Ga:YIG films were recorded in an in-plane magnetic field using a stripline connected to a Rohde & Schwarz ZVA-67 network analyzer. The details may be found in Ref. [43].

## Acknowledgements

We are grateful to A. Gopakumar for carefully reading the manuscript. We thank J. Dellith for the WD-EPMA measurements, M. Herz for the DSC measurements and R. Meyer for the technical support. We would like to thank Ferroics Inc. Betlehem, PA, USA for providing the MnZn ferrite samples. This research was supported by the Deutsche Forschungsgemeinschaft (DFG, German Research Foundation) – 271741898.

Supporting Information

# Magnetically Compensated Nanometer-thin Ga-Substituted Yttrium Iron Garnet (Ga:YIG) Films with Robust Perpendicular Magnetic Anisotropy

*Carsten Dubs\* Oleksii Surzhenko*

INNOVENT e.V. Technologieentwicklung, 07749 Jena, Germany

**Magnetic characterization by Vibrating Sample Magnetometry (VSM)**

Isothermal measurement of hysteresis loops m(H,T=Const) is a standard method for obtaining comprehensive information about key magnetic parameters of materials. Due to technological limitations, thin LPE grown Ga:YIG films cannot be separated from their GGG substrates, which typically have a thickness of 0.5 mm. These films remain free of cracks if their thickness does not exceed approximately 0.5 µm. For usual ratios $V_{GGG}/V_{Ga:YIG} \geq 10^3$, the vain contributions $m_{GGG} = \chi_{GGG}(T)HV_{GGG}$ from the GGG substrate may dominate over useful smaller signals $m_{Ga:YIG}(H,T)$, which we aim to discuss. A non-destructive method for the separation of Ga:YIG and GGG contributions from isothermal hysteresis loops at room temperature is known for conventional YIG/GGG films (see e.g., Ref.[35]). For the in-plane magnetized YIG films with $4\pi M_s$ (T=300K) ~1800 G, saturation typically occurs at small magnetic fields of a few Oersteds ($|H|>H_s$). The derivative $dm_{VSM}/dH=\chi_{GGG}(H>H_s,T)V_{GGG}=Const.$, allowed authors[35] to interpolate the GGG magnetic moment for fields $|H|\leq H_s$ and, thus, to extract $m_{YIG}(H,T)$ mathematically.

When comparing the magnetization values of Ga:YIG films with those of unsubstituted YIG films,[35] two main issues affecting the accuracy of the method must be emphasized. Firstly, the magnetic fields $|H|>H_s$ required to reach saturation magnetization in Ga:YIG films are significantly higher than for YIG films. Consequently, the useful signal $m_{Ga:YIG}(H>H_S,T=300K)$ must be mathematically subtracted from the dominating GGG signal $m_{GGG} \gg m_{Ga:YIG}$. In addition, the paramagnetic contribution $m_{GGG}(H,T) \sim HV_{GGG}/(T-\Delta)$ increases with the cooling of the sample. Here $\Delta = -2K$ is the Curie-Weiss temperature.[52] This explains why we hereafter focused on hysteresis loops at relatively high temperatures T≥300K.

The second issue is that $m_{Ga:YIG}(H>H_S,T=300K)$ in Ga:YIG films with robust PMA is relatively small. For samples in Figure 2b in the main text, saturation magnetization $4\pi M_s$ (T=300K) does not exceed 50 G, meaning their magnetic moments $m_{Ga:YIG}(H>H_S,T=300K)$ are 35 times smaller than those in YIG samples of similar geometry. As both these tendencies become more prominent near the compensation point $T_{comp}$, the hysteresis loops data $m_{Ga:YIG}(H,T \sim T_{comp})$ appear noisy, resulting in significant errors in the extracted key parameters.

**Wavelengths-Dispersive Electron Probe Micro Analysis (WD-EPMA) measurements**

Ga:YIG reference films with micrometer thickness grown on (111) GGG substrates by the LPE method were characterized by electron probe microanalysis (EPMA) (see 7. *Experimental Section* in the main text). The purpose was to confirm the correlation between saturation magnetization and the Ga content ($x_{Ga}$) in formula units (f.u.) in substituted $Y_3Fe_{5-x}Ga_xO_{12}$ films. To avoid superposition of the Ga signals from the film and substrate, the thickness of the investigated films must be in the micrometer range and the excitation energy must be as low as possible (in our case $E_0$ = 10 keV), as confirmed by Monte-Carlo simulations. For these thick films with thicknesses around 2 µm or greater (with one exception, the sample with 1.6 µm), which were grown in the same series of experiments as the nanometer-thin films in this study, the Ga content was determined and plotted as a function of saturation magnetization.

The main graph in Figure S1 shows that the correlation between the Ga concentration, as determined by WD-EPMA, and the magnetization values of the micrometer-thick films, as measured by VSM agrees well with the correlation reported for polished spheres of Ga:YIG single crystals[33,34]. However, our focus was on samples with robust perpendicular magnetic anisotropy (PMA) and those with a compensation point $T_{comp}$. Therefore, we focused on samples with a saturation magnetization at room temperature between 0 G and 200 G, corresponding to a Ga content in the range of 1.0 f.u. $\leq x_{Ga} \leq$ 1.3 f.u. (see inset in Fig. S1). All micrometer-thick films used to generate the data points in Figure S1 exhibit stress cracks, likely due to the high Ga content in the films and the resulting large lattice misfits with the GGG substrate. In contrast, the submicrometer-thin films with t ≤ 500 nm, which are discussed in the main text and listed here in Table S1, were crack-free. For submicrometer-thin films investigated in this study, quantitative analysis via WD-EPMA was not successful due to the superposition of the Ga signals of the films with the GGG substrate.

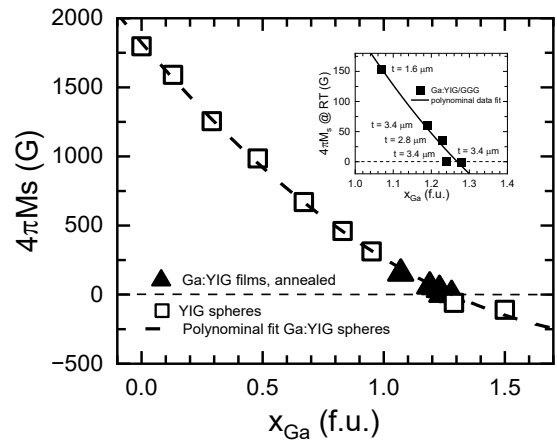

Fig. S1

Saturation magnetization $4\pi M_s$ at 300 K as a function of Ga content $x_{Ga}$ for thermally equilibrated (600°C, ≥14 h) micrometer-thick $Y_3Fe_{5-x}Ga_xO_{12}$ films grown on (111) GGG substrates (solid symbols) compared with literature data [33] for Ga:YIG spheres (open symbols). The inset shows the detailed view of the data points and their corresponding fit curve for the Ga:YIG films with 1.0 f.u. ≤ x ≤ 1.3 f.u.

Table S1

Characteristic data of Ga:YIG films grown under comparable dynamic conditions

| sample | thickness [nm] | $4\pi M_r$ [a)] [G] | $H_c$ [a)] [Oe] | $T_{comp}$ [K] | $T_{curie}$ [K] | $T_{Curie} - T_{comp}$ | $\Delta H_{FWHM}$ [a)] @6.5GHz [Oe] |
|---|---|---|---|---|---|---|---|
| A | 161 | -24 | 200 | ~345 | ≥345 | - | - |
| B | 539 | -16 | 70 | 333 | 360 | 27 | - |
| C | 200 | -15 | 90 | 335 | 362 | 27 | - |
| D | 93 | -11 | 90 | 331 | 357 | 26 | - |
| E | 205 | -10 | 120 | 325 | 357 | 32 | - |
| F | 93 | -8 | nm | 319 | 357 | 38 | - |
| G | 93 | -5 | nm | 319 | 357 | 38 | - |
| H[b)] | 865 | -4 | 100 | 311 | 367 | 56 | - |
| I | 121 | -1.3 | nm | 303 | 367 | 64 | - |
| J | 138 | 1.3 | nm | 301 | 369 | 68 | - |
| K | 130 | 5 | nm | 284 | 364 | 80 | - |
| L | 156 | 13 | 120 | 262 | 372 | 110 | - |
| M | 207 | 16 | 70 | 245 | 366 | 122 | - |
| N | 217 | 24 | 60 | 212 | 366 | 154 | - |
| O | 224 | 36 | 50 | 146 | 369 | 223 | - |
| P | 441 | 48 | 40 | 133 | 366 | 233 | - |
| Q | 232 | 55 | 50 | 98 | 365 | 267 | ~30 |
| R | 94 | 75 | 130 | - | 373 | - | 17 |
| S | 103 | 91 | 50 | - | 377 | - | 15 |
| T | 148 | 120 | 34 | - | 384 | - | 12 |
| U | 187 | 140 | 19 | - | 403 | - | 12 |
| V[c)] | 105 | 230 | 9 | - | 406 | - | 5 |
| W[c)] | 105 | 230 | 4 | - | 406 | - | 5 |
| X[c,d)] | 166 | 216 | 0.4 | - | 417 | - | 5 |

[a)] values obtained for 300 K

[b)] film shows cracks because film thickness is larger than the critical thickness for crack formation

[c)] substrate rotation rate: 100 rpm, all other samples: 200 rpm

[d)] no more rectangular hysteresis loop observed

nm = not measurable

**Stress-induced anisotropy parameter K$_\sigma$**

The reason for the constancy of the stress-induced uniaxial anisotropy parameter $K_\sigma$ in Ga:YIG films is its dependence on the in-plane stress parameter $\sigma'_\parallel$ for $\langle 111 \rangle$ oriented films and the associated magnetostriction constant $\lambda_{111}$, which can be described as follows (cf. Supporting Information of [42]),

$$K_\sigma = -\frac{3}{2} \sigma'_\parallel \lambda_{111} = -\frac{3}{2} \frac{\Delta a^\parallel}{a_f^\parallel} \frac{E}{1-\mu} \lambda_{111}.$$

While the stress parameter $\sigma'_\parallel$ increases with increasing Ga content, which is caused by the increasing in-plane lattice misfit $\Delta a^\parallel$, due to the reduction of the in-plane film lattice parameters $a_f^\parallel$, the negative magnetostriction parameter $\lambda_{111}$ decreases due to the reduction of the magnetostrictive effect caused by the reduction in the total content of tetrahedral iron ions (see e.g. Ref.[38]). This results in only minor changes in $K_\sigma$, as the Young's modulus $E$ and the Poisson's ratio $\mu$ remain nearly unchanged. Therefore, the effective perpendicular uniaxial anisotropy field $H_{2\perp}$ in Equation 2 in the main text is mainly determined by the $M_s$ value of the second term.

**Benchmarking discussion using room temperature data of Table S2**

*Column "easy axis" and "t"*
The benchmarking shows that YIG, substituted YIG or REIG films grown on different substrates exhibit robust perpendicular magnetic anisotropy only at film thicknesses between 6 nm and 100 nm (with the exception of TbIG) (compare column "t" in row "perp"), while the PMA decreases drastically with increasing film thickness (compare column "t" in row "ip"). The reason for this is that the strain relaxation due to the large lattice misfit turns the perpendicular magnetic anisotropy into the film plane with increasing film thickness. This is not the case with Ga:YIG films, which maintain their magnetic easy axis even at thicknesses up to 500 nm.

*Column "Mr/Ms"*
The degree of robustness, which means a high remanent magnetization at zero field, is not achieved with < 90% for some of the films, not even for films with thicknesses between 10 nm and 30 nm. For Ga:YIG films, this is the rule even for micrometer-thick films.

*Column "H$_c$"*
The coercive field strength H$_c$ is in the range between ~0 Oe and 100 Oe for most samples, with the exception of reports for the rare-earth iron garnets TbIG and TmIG with up to ~1500 Oe. However, no damping data and large FMR linewidths ΔH$_{p-p}$ of up to 540 Oe and Gilbert damping coefficients α between 0.01 and 0.017 are reported for these films, which makes them difficult to use for devices.

*Column "4πM$_s$"*
Most of the iron garnet films exhibit high saturation magnetizations, with the exception of a few reports of rare-earth garnets such as TbIG and GdIG and single examples of substituted Bi:YIG and TmIG. In the case of Ga:YIG, the saturation magnetization can be continuously varied between 1800 G and 0 G due to the dilution of the iron ions by Ga ions on the different magnetic sublattices sites. The Table S2 gives examples with 4πM$_s$ values below 200 G, which are significantly lower than previously reported. Therefore, shape anisotropy can be easily overcome by uniaxial anisotropy contributions, leading to robust PMA behavior (large $4\pi M_{eff}$ values), even for epitaxial films several hundred nanometers thick.

*Column "$4\pi M_{eff}$"*

The effective magnetization, which defines the magnitude of the required magnetic field that must be provided to rotate the magnetization into the plane, varies between -290 G and -1400 G for most of the samples, except for TbIG and Ga:YIG samples. However, it should be noted that in contrast to the Ga:YIG samples with thicknesses beyond 100 nm, high $4\pi M_{eff}$ values are only reported for very thin films of TmIG with thicknesses of 6 nm (-1240 G) or for YIG of 8 nm (-1430 G). Unfortunately, no FMR data were published for TmIG and TbIG so that their suitability for devices with low damping cannot be verified. In contrast, the absolute damping determined for the Ga:YIG films are the lowest of all the PMA films listed in the Table S2.

*Column "$\Delta H^{\parallel}_{p-p}$" and "$\Delta H^{\perp}_{p-p}$"*

The FMR linewidth $\Delta H_{p-p}$, which is a measure of the absolute damping values at defined frequencies (sum of the intrinsic and inhomogeneous damping contributions), are listed for in-plane and perpendicular oriented magnetic bias fields and confirm that the Ga:YIG samples show the narrowest linewidths of all listed samples. Thin films of pure YIG and the Mn- or some of the Bi-substituted YIG films between 6 nm and 30 nm thickness show an order of magnitude higher linewidth compared to Ga:YIG films [19,20], while all other films exhibit up to two orders of magnitude higher values. In contrast, narrow linewidths can also be obtained for Ga:YIG films with thicknesses of about 200 nm and $4\pi M_{eff}$ values as large as -2300 G.

*Column "$\alpha$"*

For the Gilbert damping coefficient $\alpha$ (a measure of the intrinsic damping contributions), which can be determined from the frequency dependence of the FMR linewidths, the best values were obtained for pure YIG with a thickness of 8 nm and Ga:YIG with a thickness of 187 nm, while all other films exhibit larger intrinsic damping contributions. For the Ga:YIG film with the largest PMA of $4\pi M_{eff} \sim -2300$ G, $\alpha$ is increased, but significantly lower than for most substituted and rare-earth iron garnets.

*Column "spin wave decay length"*

However another major advantage of Ga:YIG films is the about one order of magnitude larger spin wave decay length resulting from the intrinsic exchange stiffness, which is three times larger than for pure or substituted YIG films in this table. Even if this value was not determined for the Ga:YIG films grown in this work, it can be assumed that they should have comparable values.

Table S2    Benchmarking of (111) YIG or substituted YIG films with robust perpendicular magnetic anisotropy at room temperature

| film | substrate[a] | easy axis ip[b] | t [nm] | $M_r/M_s$ [%] | $H_c$ [Oe] | $4\pi M_s$ [G] | $4\pi M_{eff}$ [G] | $\Delta H^{\|}_{p-p}$ (GHz) [Oe] | $\Delta H^{\perp}_{p-p}$ (GHz) [Oe] | $\alpha \times 10^{-4}$ | spin wave decay length [μm] | remarks | reference/ technique |
|---|---|---|---|---|---|---|---|---|---|---|---|---|---|
| YIG | GSGG | perp | 8 | ~97 | 8 | 1430 | -500/ | - | 20 (7) | 4 | - | GSGG substrates used from different sources | [1] sputtering |
| | | perp | 8 | - | - | - | -1400 | - | - | - | - | | |
| | | ip | ≥10 | ≤64 | - | - | - | - | - | - | - | | |
| YIG | GYSGG | perp | 10 | ~80 | 2 | 1600 | -300 | - | - | - | - | YIG films on YSSG substrates did not result in a robust PMA, even for films of 10nm thickness | [2] sputtering |
| | | ip | >15 | - | - | - | - | - | - | - | - | | |
| | GSGG | perp | 10 | ~100 | 50 | - | -600 | - | - | - | - | | |
| | | ip | >10 | - | - | - | - | - | - | - | - | | |
| Mn:YIG | GGG | perp | 30 | ~40 | ~0 | 1570 | -530 | - | 20 (4) | 80 | 3.7 | Kittel data fitting from 2 - 4 GHz data only; $Mn_{1.12}$ | [3] PLD |
| Bi:YIG | sGGG | perp | 20 | ~100 | ~5 | 850 | -910 | - | 40 (7) | 54 | 0 | the propagation of spin wave was not observed | [11] PLD |
| | | ip | >60 | ≤30 | - | - | - | - | - | - | - | | |
| $Bi_1Y_2IG$ | sGGG | perp | 18 | ~75 | 3 | 1440 | - | - | 20 (9) | 7 | - | shape overcomes the strain anisotropy at t=20nm | [12] PLD |
| | | ip | ≥25 | - | - | - | - | - | 20 (9) | - | - | | |
| $Bi_1Y_2IG$ | sGGG | perp | 30 | ~100 | ~7 | 1720 | -290 | 180 (8) | 900 (8) | ? | 3.7 [14] | α not comparable, as only determined for θ=30.5° | [13]/[14] PLD |
| | | ip | > 32 | - | - | - | - | - | - | - | - | | |
| TbIG | GGG | perp | 100 | <100 | 1500 | 220 | -8800[c] | - | - | - | - | - | [4] PLD |
| TmIG | GGG | perp | 6 | ~100 | 300 | 1260 | -1240[d] | - | - | - | - | - | [5] PLD |
| TmIG | GGG | perp | 15 | - | - | 1260 | -980 | 140 (9) | 220 (9) | - | - | specified half linewidths ΔH were multiplied by 2 | [6] sputtering |
| | | perp | 30 | - | - | 1260 | -800 | 130 (9) | 160 (9) | 150 | - | | |
| | | perp | 60 | - | - | 1260 | -380 | 120 (9) | 140 (9) | 150 | - | | |
| TmIG | sGGG | perp | 50 | ~100 | 230 | 880 | -710 | 360 (9) | 540 (9) | 170 | - | α is an averaged value for films of t = 20-200nm | [7] PLD |
| | | ip | >70 | - | - | - | - | - | - | - | - | | |
| GdIG | sGGG | perp | 35 | - | - | ~630 | - | 90 (6) | - | 100 | - | $T_{comp}$=200K | [8] PLD |
| Ga:YIG | GGG | perp | 45 | 100 | 40 | 190 | -700 | 4 (7) | - | 10 | 30[e] | - | [19] LPE |
| Ga:YIG | GGG | perp | 59 | - | - | 200 | -700 | 6 (7) | - | 6 | 30[f] | - | [20] LPE |
| Ga:YIG | GGG | perp | 187 | 99 | 2 | 130 | -1500[g] | 7 (7) | - | 4[g] | - | #U; no $T_{comp}$ | this work LPE |
| | | perp | 232 | ~100 | 5 | 60 | -2300[g] | 16 (7) | - | 21[g] | - | #Q; $T_{comp}$= 73K, | |
| | | perp | 217 | 80 | 60 | 24 | - | - | - | - | - | #N; $T_{comp}$= 212K | |
| | | perp | 539 | 93 | 70 | -160 | - | - | - | - | - | #B; $T_{comp}$= 360K | |

[a] GSGG: $Gd_3(Sc_2Ga_3)O_{12}$, GYSGG: $(Gd_{0.63}Y_{2.37})(Sc_2Ga_3)O_{12}$, $Y_3(Sc_2Ga_3)O_{12}$, GGG: $Gd_3Ga_5O_{12}$, sGGG: $Gd_3(CaMgZrGa_3)O_{12}$, [b] (ip) degradation of the robust PMA starts for films with further increasing thickness, [c] $4\pi M_{eff}$ from magneto-transport data, [d] from Hall resistance data [e] from all-electrical spectroscopy data, [f] from BLS spectroscopy data, [g] K. Lenz et al., poster presentation at Magnonics 2023, 30.07.-03.08., Le Touquet - Paris - Plage, France; Publication in preparation